# Hybrid Recommendation System using Graph Neural Network and BERT Embeddings

**Shashidhar Reddy Javaji**
University of Massachusetts Amherst
Amherst, MA
`sjavaji@umass.edu`

**Krutika Sarode**
University of Massachusetts Amherst
Amherst, MA
`ksarode@umass.edu`

## Abstract

Recommender systems have emerged as a crucial component of the modern web ecosystem. The effectiveness and accuracy of such systems are critical for providing users with personalized recommendations that meet their specific interests and needs. In this paper, we introduce a novel model that utilizes a Graph Neural Network (GNN) in conjunction with sentence transformer embeddings to predict anime recommendations for different users. Our model employs the task of link prediction to create a recommendation system that considers both the features of anime and user interactions with different anime. The hybridization of the GNN and transformer embeddings enables us to capture both inter-level and intra-level features of anime data. Our model not only recommends anime to users but also predicts the rating a specific user would give to an anime. We utilize the Graph-SAGE network for model building and weighted root mean square error (RMSE) to evaluate the performance of the model. Our approach has the potential to significantly enhance the accuracy and effectiveness of anime recommendation systems and can be extended to other domains that require personalized recommendations.

## 1 Introduction

Recommendation systems are algorithms that suggest items to users based on their past behavior. They are used in a variety of applications, such as online shopping, music streaming, and social media. There are two main types of recommendation systems: collaborative filtering and content-based filtering. Collaborative filtering systems recommend items to users based on the ratings or preferences of other users. For example, if you have rated a number of movies on Netflix, the collaborative filtering system will recommend other movies that other users with similar ratings have also enjoyed.

Content-based filtering systems recommend items to users based on the content of the items themselves. For example, if you have listened to a number of songs by a particular artist, the content-based filtering system will recommend other songs by the same artist. In recent years, there has been a trend towards using hybrid recommendation systems that combine the strengths of collaborative filtering and content-based filtering. These systems can provide more accurate recommendations than either type of system on its own.

There are a number of different ways to build recommendation systems. One common approach is to use machine learning algorithms. Machine learning algorithms can be trained on large datasets of user ratings or preferences to learn how to predict which items a user will like. Another approach to building recommendation systems is to use artificial intelligence (AI) techniques. AI techniques, such as deep learning, can be used to create more complex and powerful recommendation systems. Recommendation systems have become an integral part of our daily lives, aiding us in making informed decisions about the products and services we use. The success of these systems can be attributed to their ability to filter and personalize vast amounts of information, making it easier for users to find relevant and useful items. However, the increasing complexity and heterogeneity of data have made it challenging to develop accurate and efficient recommendation systems.

In recent years, graph neural networks (GNNs) have emerged as a promising solution to this problem, allowing us to incorporate relational data into our recommendation models. GNNs can effectively capture the inherent structure and dependencies in the data, enabling us to make more accurate and personalized recommendations. Graph Neural Networks (GNNs) have emerged as a powerful approach to solving problems in the domain of recommendation systems. Recommendation systems aim to recommend items to users that are relevant and useful to them, based on their past behavior and preferences. GNNs can help in creating better recommendations by modeling the complex relationships between users and items in a graph-based representation. One of the key challenges in recommendation systems is the sparsity of the data. In many cases, users may have only interacted with a small subset of items, and the available data may not be sufficient to learn accurate models. GNNs can help address this challenge by leveraging the graph structure of the data to propagate information from observed to unobserved nodes.

GNNs can be used in both content-based and collaborative filtering approaches to the recommendation. In a content-based approach, GNNs can be used to model the features of the items and users and create recommendations based on the similarity between their embeddings. In a collaborative filtering approach, GNNs can be used to model the interactions between users and items in a graph, and create recommendations based on the relationships between the nodes.

One of the popular approaches for GNN-based recommendation is GraphSAGE. GraphSAGE is a variant of GNN that aggregates information from neighboring nodes to generate node embeddings. In GraphSAGE, each node is assigned an initial feature vector, and these features are updated iteratively by aggregating information from the node's neighbors. The aggregated features are then passed through a neural network layer to generate a new embedding for the node. In the context of recommendation, GraphSAGE can be used to generate embeddings for both users and items. The model can be trained to predict the likelihood of a user interacting with an item, based on the embeddings of the user and item. The learned embeddings can then be used to generate recommendations for users.

To improve the performance of the recommendation system, additional features can be incorporated into the model. For example, in the case of movie recommendations, features such as the genre and the synopsis of the movie can be used to augment the embeddings of the movies. Similarly, features such as the age and gender of the user can be used to augment the embeddings of the users. Overall, GNNs have shown great promise in the domain of recommendation systems and can help in creating more accurate and personalized recommendations for users. With the availability of large amounts of data and the increasing interest in personalized recommendations

The rest of the paper is organized as follows: Section 2 provides a brief overview of related work. Section 3 describes the dataset and the pre-processing steps used to prepare the data. Section 4 presents the proposed model in detail. Section 5 presents the experimental setup and results. Finally, Section 6 concludes the paper with a summary of the contributions and directions for future work

## 2  Related Work

Recommender systems have been widely used to provide personalized recommendations to users. Collaborative filtering (CF) is a popular technique that utilizes users' past behavior to make recommendations. Matrix factorization, a type of CF algorithm, decomposes the user-item interaction matrix into two lower-dimensional matrices to represent users and items. The regularization weights of the latent factors can be assigned based on items' popularity and users' activeness, which can improve the prediction results of the matrix factorization technique. [4]

The paper on graph neural networks in recommender systems provides a survey of various graph-based techniques for recommender systems, including GCNs, GATs, and GAEs. The paper discusses how these techniques can be used to handle cold-start problems, incorporate side information, and enhance recommendation accuracy. [5] Graph-based models have become increasingly popular in recent years for their ability to handle complex interactions between users and items. The linear residual graph convolutional network approach for CF-based recommender systems revisits GCNs in CF models and shows that removing non-linearities can enhance recommendation performance. The proposed model uses a residual network structure that is specifically designed for CF with user-item interaction modeling, which alleviates the over-smoothing problem in graph convolution aggregation operation with sparse data. [3]

The graph-based hybrid recommendation system (GHRS) combines content-based and collaborative filtering approaches to extract new features based on users' ratings, demographic, and location information. These features are then used for clustering users, which improves recommendation accuracy and dominates other methods' performance in the cold-start problem. The experimental results on the MovieLens dataset show that the proposed algorithm outperforms many existing recommendation algorithms on recommendation accuracy. [1]

Inductive matrix completion is another popular approach to building recommender systems that can handle the cold-start problem. The paper on learning to transfer graph embeddings for inductive graph-based recommendation proposes a transfer learning framework for personalized video highlight recommendation. The proposed framework is composed of two parts: a graph neural network that exploits the higher-order proximity between users and segments to alleviate the user cold-start problem and an item embedding transfer network that approximates the learned item embeddings from graph neural networks. [2]

Matrix factorization, specifically, is a widely used technique in recommender systems that utilizes users' past behavior, such as ratings or purchases, to make recommendations. One of the most popular CF algorithms is matrix factorization, which decomposes the user-item interaction matrix into the product of two lower dimensionality rectangular matrices, user and item embeddings, that represent users and items in a lower-dimensional space. The regularization weights of the latent factors can be assigned based on items' popularity and users' activeness, which can improve the prediction results of the matrix factorization technique. The paper on matrix factorization techniques for recommender systems provides a foundational understanding of collaborative filtering and matrix factorization for building recommender systems. [4]

In summary, the related papers cover various techniques for building recommender systems, including matrix factorization, graph-based models, inductive matrix completion, and transfer learning. These papers provide further insights into the use of these techniques in recommender systems and how they can be used to handle cold-start problems, incorporate side information, and enhance recommendation accuracy.

## 3 Dataset

The Anime Recommendation Database 2020 is a dataset available on Kaggle, containing information about anime and user interactions from the website MyAnimeList. The dataset was created by scraping the website and contains recommendation data from 320,000 users and 16,000 animes.

The dataset is comprised of two main tables: the anime table and the rating table. The anime table contains information about each anime, including its ID, name, genre, type, episodes, and synopsis. The genre field is a list of genres associated with anime, such as "Action", "Comedy", "Drama", and "Fantasy". The type field indicates whether the anime is a TV series, movie, OVA, or other formats. The episodes field indicates the number of episodes in the series. The synopsis field provides a brief description of the anime's plot.

The rating table contains information about user interactions with the animes, including the user ID, the anime ID, and the user's rating for the anime on a scale of 1 to 10. The dataset also includes a timestamp field indicating the time when the user rated the anime.

The dataset contains a total of 78,460,895 user-anime interactions, with an average of 4.9 ratings per user. The most popular anime in the dataset is "Death Note", with over 150,000 ratings. The dataset is useful for building recommendation systems for anime, as it contains information about both the animes and user preferences.

### 3.1 Preprocessing

The dataset used in this research consists of two primary data sources: the "anime with synopsis" and "rating complete" files, which were merged to obtain relevant columns for the model. Specifically, the dataset includes anime id, user id, synopsis, genres, and rating. Prior to analysis, the dataset underwent a preprocessing step which involved data cleaning to remove rows with null values in any column. One hot encoding was also applied to the genres column in order to transform the categorical variable into a numerical format suitable for analysis.

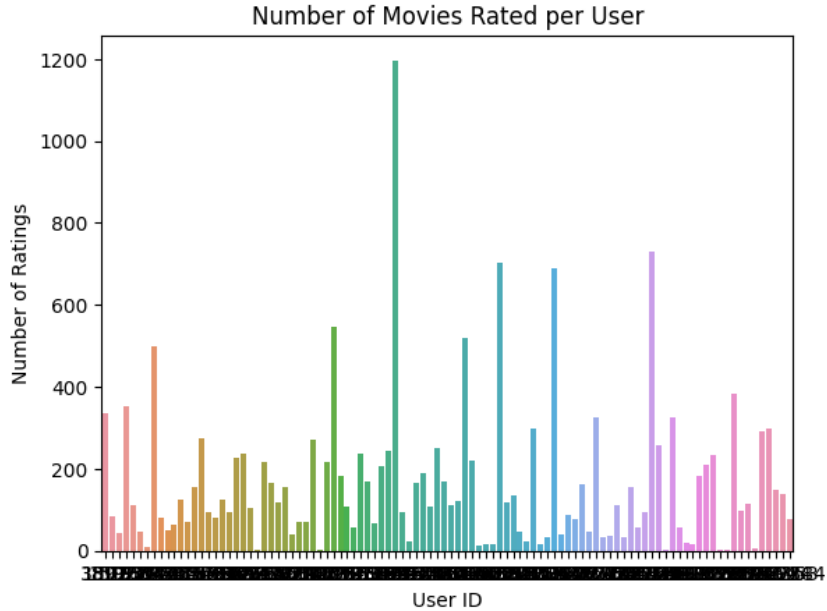

Figure 1: Bar graph of ratings given by each users

Furthermore, two dictionaries were created to map the user id's and anime id's in the dataset. These dictionaries were used to facilitate the analysis and interpretation of the data. Overall, the resulting dataset is suitable for use in conducting research on anime recommendation systems, and provides a robust foundation for the development and evaluation of machine learning algorithms for this purpose.

We created three classes: SequenceEncoder, IdentityEncoder, and GenresEncoder, which encode different types of data into PyTorch tensors. These classes are used to load and process node and edge data for a graph-based recommendation system. The SequenceEncoder class encodes text data using the SentenceTransformer model. The input data is a Pandas dataframe, and the output is a PyTorch tensor that represents the sentence embeddings. The IdentityEncoder class converts raw column values to PyTorch tensors, and the GenresEncoder class encodes genre information from the raw data. The load node csv function uses these encoders to process the node data, concatenating the resulting tensors into a single tensor.

The load edge csv function loads edge data and generates labels for each edge. It takes two arguments, ratings user id and ratings movie id, which are the user and movie IDs for each rating. It then generates edge labels by looking up the corresponding ratings from a dictionary user anime rating and returns a PyTorch tensor containing the edge labels. Overall, the code shows how the dataset is preprocessed before being fed into the graph-based recommendation system. The SequenceEncoder, IdentityEncoder, and GenresEncoder classes are used to encode different types of data into PyTorch tensors, which are then concatenated into a single tensor using the load node csv function. The load edge csv function loads edge data and generates labels for each edge, completing the dataset preprocessing pipeline.

## 4  Proposed Methodology

In an anime recommendation system, the features used for node creation can have a significant impact on the performance of the system. One common approach is to use genres as the features for each anime. Genres are categorical variables that can be one-hot encoded and used to represent the anime's content. This approach is straightforward and easy to implement, but it has some limitations.

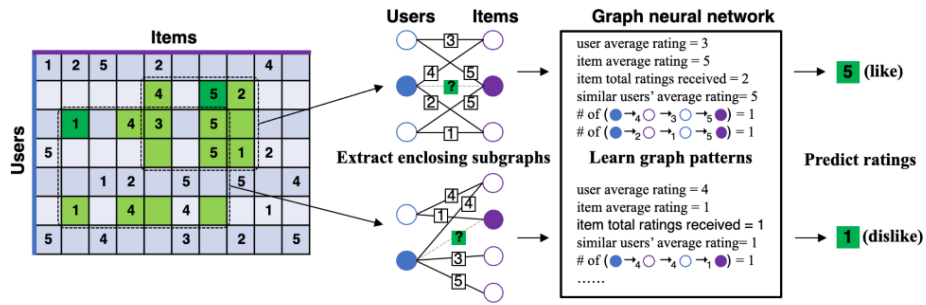

Figure 2: Architecture of the model

One limitation is that genres alone may not capture the complexity and nuances of the anime. For example, two anime could have the same genres, but one could be a comedy with a light-hearted tone while the other could be a dark psychological thriller. In this case, relying solely on genres may not differentiate between the two anime and could lead to poor recommendations.

To overcome this limitation, we can combine the genres with the sentence embeddings of the synopsis. The synopsis is a brief summary of the anime's plot, and it can provide additional information about the anime's content and style. By using sentence embeddings, we can capture the meaning and context of the synopsis, which can help to differentiate between anime with similar genres.

To do this, we first preprocess the synopsis by removing stop words, punctuation, and other irrelevant information. We then use a pre-trained sentence embedding model such as BERT or GloVe to generate embeddings for each sentence in the synopsis. We can then average these embeddings to obtain a single embedding for the entire synopsis. We can then concatenate the one-hot encoded genres with the synopsis embedding to create a feature vector for each anime. This feature vector captures both the categorical information about the anime's genres and the semantic information about the anime's content and style. Once we have the feature vectors for each anime, we can use them to create nodes in the graph. We can then use graph neural networks (GNNs) to learn the representations of these nodes and generate recommendations based on the learned representations. Compared to using genres alone, combining genres with the synopsis embeddings can lead to more accurate and personalized recommendations. This approach can capture the complex and nuanced content of the anime and provide better differentiation between anime with similar genres. Additionally, this approach can be extended to incorporate other textual features such as reviews or user feedback, which can further improve the recommendations.

The Model class inherits from the PyTorch Module class, which provides a convenient way to define a neural network model. The __init__ method defines the components of the model and initializes their parameters. The forward method defines the computation that will be performed by the model when it is run on input data. The GNNEncoder class is a custom implementation of a GNN encoder that takes as input a set of node features and edge connections and outputs a set of node embeddings. The hidden_channels argument specifies the dimensionality of the node embeddings. The GNNEncoder class is defined in a separate file and is not shown in the code snippet provided.

```
HeteroData(
  user={ x=[100, 100] },
  anime={ x=[3534, 427] },
  (user, rates, anime)={
    edge_index=[2, 16143],
    edge_label=[16143]
  },
  (anime, rev_rates, user)={ edge_index=[2, 16143] }
)
```

: HeteroData Structure

The encoder attribute of the Model class is an instance of the GNNEncoder class. It takes the hidden_channels argument as input and is initialized with the same dimensionality for both the input and output features.

The to_hetero function is a utility function that converts the GNNEncoder object to a heterogeneous GNN. The data.metadata() argument specifies the schema of the heterogeneous graph, which includes information about the node types, edge types, and features of the graph. The aggr argument specifies the type of aggregation to be used when combining information from different node types. The EdgeDecoder class is a custom implementation of an edge decoder that takes as input a set of node embeddings and a set of edge connections and outputs a set of edge predictions. The hidden_channels argument specifies the dimensionality of the node embeddings. In the GNNEncoder class, the GraphSAGE implementation is achieved by using the SAGEConv module from PyTorch Geometric library. The SAGEConv module implements the GraphSAGE convolutional operator, which aggregates the feature vectors of a node and its neighbors using a graph convolutional operation.

The decoder attribute of the Model class is an instance of the EdgeDecoder class. It takes the hidden_channels argument as input and is initialized with the same dimensionality for both the input and output features. The forward method takes as input a dictionary of node features, a dictionary of edge connections, and a set of edge labels. The x_dict argument is a dictionary of PyTorch tensors representing the node features for each node type. The edge_index_dict argument is a dictionary of PyTorch tensors representing the edge connections for each edge type. The edge_label_index argument is a PyTorch tensor representing the edge labels.The forward method of the GNNEncoder class first applies a GraphSAGE layer to the input node features using the SAGEConv module. This layer aggregates the feature vectors of each node and its neighbors using a graph convolutional operation. The resulting feature vectors are then normalized and passed through a ReLU activation function.

The forward method first passes the input data through the encoder to obtain a set of node embeddings, represented as a dictionary of PyTorch tensors. It then passes these node embeddings and the edge labels through the decoder to obtain a set of predicted edge labels. In summary, the model architecture consists of a GNN encoder that takes as input node features and edge connections, a heterophily operator that converts the GNN encoder to a heterogeneous GNN, and an edge decoder that takes as input node embeddings and edge connections and outputs a set of predicted edge labels. The model is designed for semi-supervised learning on heterogeneous graphs and can handle multiple node and edge types with different feature representations.

In the context of graph neural networks (GNNs), the heterophily operator is a mechanism used to combine information from nodes of different types in a heterogeneous graph. In a heterogeneous graph, nodes can have different types, which correspond to different features or attributes. For example, in a citation network, nodes can represent papers, authors, or conferences, and each node type can have different attributes such as publication year, paper topic, or author affiliation. To capture such heterogeneity, GNNs use different weight matrices for each node type, allowing the model to learn different representations for nodes of different types.

In the GNNEncoder class, the GraphSAGE implementation is achieved by using the SAGEConv module from the PyTorch Geometric library. The SAGEConv module implements the GraphSAGE convolutional operator, which aggregates the feature vectors of a node and its neighbors using a graph convolutional operation. The GNNEncoder class takes two arguments: the number of input feature dimensions and the number of output feature dimensions. The forward method of this class applies two GraphSAGE layers to the input node features to generate the output node features. The forward method of the GNNEncoder class first applies a GraphSAGE layer to the input node features using the SAGEConv module. This layer aggregates the feature vectors of each node and its neighbors using a graph convolutional operation. The resulting feature vectors are then normalized and passed through a ReLU activation function. The output of the first GraphSAGE layer is then passed through a second GraphSAGE layer in a similar fashion. Finally, the resulting output features are returned as the output of the forward method of the GNNEncoder class. Overall, the GNNEncoder class implements a GraphSAGE-based neural network architecture for learning node representations in a graph by aggregating neighborhood information of each node in the graph

# 5 Evaluation and Results

The process of evaluation is as follows; this model is evaluated using Root Mean square Error(RMSE), the model is used to get the ratings between a given user and certain anime which the user haven't watched before, all such links are predicted with certain weight, so given a user we get the ratings for different anime in the list which they haven't watched yet, after this the predicted ratings along with the anime are taken for particular user and then the list is sorted according to the rating predicted, we get the list of anime with highest to lowest rated for the anime that would be given by the user if watched as predicted by the model, top 10 anime of this list are taken and are recommended to that user as the anime recommendation that the user can watch. The evaluation is done by using the test set where we have the ratings that are given by the user for different anime, these are not shown at the training time, trained model is used to predict the rating and then evaluate it with the ground truth labels, using RMSE we check how close the model is able to predict the values for the given graph.

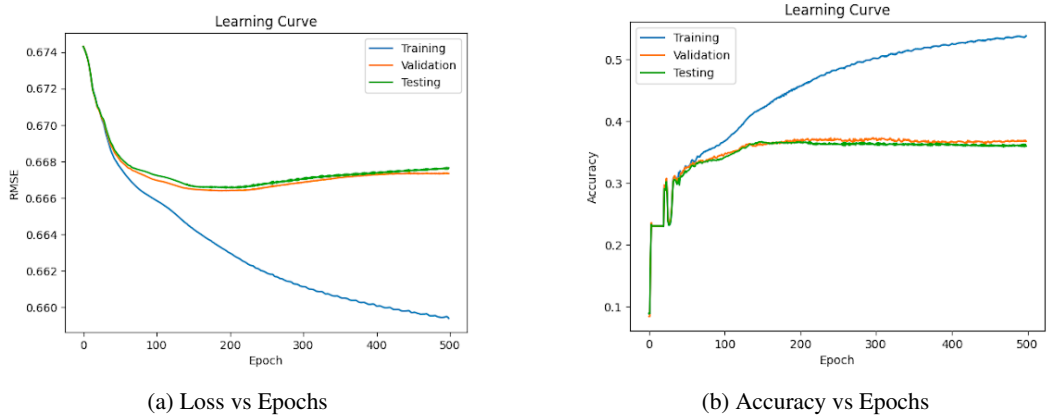

(a) Loss vs Epochs        (b) Accuracy vs Epochs

Figure 3: Results

**Recomendation for user: 415** ['Pokemon Movie 14 White: Victini to Kuroki Eiyuu Zekrom', 'Tsuki no Sango', 'Charlotte', 'Tanaka-kun wa Kyou mo Kedaruge', 'Iblard Jikan', 'Teekyuu', 'Tenshi Nanka ja Nai', 'No Game No Life: Zero', 'Puchitto Gargantia', 'Pokemon: Senritsu no Mirage Pokemon']

**Recomendation for user: 30** ['Jormungand', 'Hanayamata', 'BlackRock Shooter (OVA)', 'Mahou Shoujo Ore', 'Selector Infected WIXOSS', 'Kara no Kyoukai 6: Boukyaku Rokuon', 'Claymore', 'Kamigami no Asobi', 'Zettai Bouei Leviathan', 'Kakegurui']

**Recomendation for user: 189** ['Zero no Tsukaima F', 'Mahouka Koukou no Rettousei Movie: Hoshi wo Yobu Shoujo', 'Kami-tachi ni Hirowareta Otoko', 'Sunohara-sou no Kanrinin-san', 'Dragon Ball GT', 'Seishun Buta Yarou wa Bunny Girl Senpai no Yume wo Minai', 'Tamako Market', 'School Days', 'Kono Bijutsubu ni wa Mondai ga Aru!', 'Re:Zero kara Hajimeru Isekai Seikatsu 2nd Season']

**Recomendation for user: 298** ['Mobile Suit Gundam 00', 'Bannou Bunka Neko-Musume DASH!', 'Fullmetal Alchemist: Premium Collection', 'Naruto Movie 2: Dai Gekitotsu! Maboroshi no Chiteiiseki Dattebayo!', 'Issho ni Training: Training with Hinako', 'Doraemon', 'School Rumble', 'Golden Boy', 'Rurouni Kenshin: Meiji Kenkaku Romantan - Tsuioku-hen', 'Death Note: Rewrite']

The results of our experiment are presented in this section. The model was trained and tested on a dataset consisting of 800 users. The following are the results of our experiment:

**Train loss:** 0.659

**Test Loss:** 0.667

**Train Accuracy:** 0.52

**Test Accuracy:** 0.37

# 6 Conclusion and Future Work

The results show that the model achieved a higher accuracy on the training data (52%) compared to the testing data (37%). The loss values for both the training and testing data are relatively high, indicating that the model may not be performing optimally. Though the accuracy is not that high, but the model is working and giving good results with very less amount of data, the compute resource required to run for large amount of data is very high The future plans of the model would be to try with more nodes which can be made into features and then make edges between the user and their features as well as the anime and the features of the animes. Also the other side of future work would to try on more data which would be possible with more compute resources. There is also possibility of trying more types of GNN's other than Graph SAGE network for the training process of the GNN.